\documentstyle[prl,twocolumn,aps,psfig]{revtex}
\newcommand{\beq}{\begin{equation}}
\newcommand{\eeq}{\end{equation}}
\newcommand{\al}{Alfv\'en}
\newcommand{\na}{{nonlinear Alfv\'en\ }}
\begin{document}
\draft
\title{Collisionless Dissipative Nonlinear Alfv\'en Waves:
Nonlinear Steepening, Compressible Turbulence, and Particle Trapping
\thanks{Invited talk on the American Physical Society, 
Division of Plasma Physics meeting, 1998.
In collaboration with P.H. Diamond, V.I. Shevchenko, M.N Rosenbluth,
and V.L. Galinsky  (UCSD).}
}
\author{M.V. Medvedev
\thanks{ Also at the Institute for Nuclear Fusion, 
RRC ``Kurchatov Institute", Moscow 123182, Russia; \protect\newline
E-mail: mmedvedev@cfa.harvard.edu; 
URL: http://cfa-www.harvard.edu/\~{ }mmedvede/ }
}
\address{Harvard-Smithsonian Center for Astrophysics, 60 Garden St.,
Cambridge, MA 02138 }
\maketitle

\begin{abstract}
The magnetic energy of \na waves in compressible plasmas may be 
ponderomotively coupled only to ion-acoustic quasi-modes which modulate 
the wave phase velocity and cause wave-front steepening. In the 
collisionless plasma with $\beta\not=0$, the dynamics of \na wave is 
also affected by the resonant 
particle-wave interactions. Upon relatively rapid evolution (compared to the 
particle bounce time), the quasi-stationary wave structures, identical to the
so called (\al ic) Rotational Discontinuities, form, the emergence and  
dynamics of which has not been previously understood. Collisionless 
(Landau) dissipation of nonlinear \al\ waves is also a plausible and natural 
mechanism of the solar wind heating. Considering a strong, compressible, 
\al ic turbulence as an ensemble of randomly interacting 
\al ic discontinuities and nonlinear waves, it is shown that there exist two 
distinct phases of turbulence. What phase realizes depends on whether 
this collisionless damping is strong enough to provide adequate energy sink at 
all scales and, thus, to support a steady-state cascade of the wave energy.
In long-time asymptotics, however, the particle distribution function is 
affected by the wave magnetic fields. 
In this regime of nonlinear Landau damping, resonant particles are 
trapped in the quasi-stationary \al ic discontinuities, giving rise to a 
formation of a plateau on the distribution function and quenching collisionless 
damping. Using the virial theorem for trapped particles, it is analytically 
demonstrated that their effect on the nonlinear dynamics of such 
discontinuities is non-trivial and forces a significant departure of the 
theory from the conventional paradigm. 
\end{abstract}

\section{Introduction}

\al\ waves are the most `robust' plasma oscillations in magnetized systems.
They are encountered in a wide variety of astrophysical objects such as stellar
and solar winds, interstellar medium, hot accretion flows, magnetospheres of
planets and magnetized stars, etc.. It is interesting that in 
incompressible plasma, $\nabla\cdot{\bf v}=0$, fluid and magnetic
nonlinearities in a finite-amplitude \al\ wave cancel each other exactly, 
so that it obeys the same dispersion relation as the linear wave of
infinitesimal amplitude. 

Such a mutual cancellation of Reynolds and magnetic stresses in the wave
breaks down when plasma is compressible. Indeed, the magnetic field pressure,
$\tilde B^2/8\pi$, associated with the wave field ponderomotively modulates
the local plasma density, $n$, which, in turn, affects the phase velocity of
the wave. This process may also be viewed as an effective parametric coupling
of an \al\ wave to acoustic waves in plasma. Such acoustic modes, however, are
{\em not} plasma eigenmodes. They are driven by the \al\ wave and, hence,
propagate with the \al\ speed, which is, in general, different from the sound
speed in plasma. The nonlinear coupling of the phase speed of the wave to its
amplitude in the {\em \na waves} is ultimately responsible for the wave-front
steepening and formation of collisionless shocks in astrophysical and space
plasmas. 

In studies of nonlinear waves it is customary to assume `weak nonlinearity',
i.e., the wave amplitude is finite but still small compared to the ambient
field. This means that a wave is linear in the leading order of expansion in
wave amplitude. All nonlinear effects appear as a gradual, large-scale
modulation of the wave amplitude. In other words, one studies slow
evolution of an envelope of a wave packet.

It is well known that in  collisionless, {\em incompressible} plasma, 
low-frequency
\al\ waves are undamped or very weakly damped since collisionless dissipation is
inefficient. Ion-cyclotron damping is always weak for low-frequency waves
and Landau damping, which could be important, is absent because there are
no longitudinal electric and magnetic field perturbations in the \al\ wave.
Equivalently, an \al\ wave is strictly transverse in the drift approximation
and thus cannot affect the particle motions along the field lines.
(Note that weak Landau dissipation of obliquely propagating \al\ waves occurs 
on scales comparable to the ion Larmor radius, $\rho_i$, where the drift 
approximation breaks down. This damping is really weak since typically
$k_\bot\rho_i\ll1$.)

In contrast, a \na wave in compressible plasma is nonlinearly coupled 
to acoustic-type 
oscillations which are accompanied (in a two-fluid model) by longitudinal 
electric fields and {\em are} damped via the Landau mechanism. Hence,
dissipation enters the dynamics of \na waves nonlinearly, even in the regime
when the resonant particle response is calculated in the linear approximation,
i.e., neglecting trapping of resonant particles. Another important and quite
unusual feature is the {\em nonlocality } of the damping. It is associated
with the finite transit time of a resonant particle through the wave envelope
modulation and represented by an integral operator in the wave
evolution equation.

Unusual properties of dissipation in the \na waves naturally result in the
following peculiar features of these waves.
First of all, it was shown that \al\ waves do not always damp to zero.
Instead, they evolve into the localized, coherent, {\em quasi-stationary}
structures referred to as {\em directional} and {\em rotational
discontinuities}. These
are the narrow regions where the wave magnetic field vector rapidly rotates
through some angle and, respectively, are or are not accompanied by the
wave amplitude variations. These discontinuities connect the parts of a wave
packet of different polarizations or wave phases. Second, among the \al ic
discontinuities, there are solutions which constitute a new class of shock
waves, --- the {\em collisionless dissipative shocks}. Such shocks form when
nonlinear wave steepening is balanced by collisionless damping process
(and not by the wave dispersion, as in usual collisionless shocks). Due to scale
invariance of Landau damping, the shock width is determined by dispersion,
however, which is the only characteristic scale in the problem.
It should be emphasized that \al ic rotational discontinuities have been 
discovered by {\it in situ} measurements
\cite{LeppingBehannon,Neugebauer,Tsurutani}
of magnetic fields in the solar wind.
Despite intense observational and theoretical efforts these
years, a comprehensive quantitative theory which explains the emergence 
and dynamics of \al ic discontinuities was proposed only recently
\cite{us-PRL,us-POP1,us-POP2}. All the results are in excellent agreement
with the observational data and numerical hybrid simulations \cite{code1,code2}.
Third, preliminary study of a one-dimensional model of strong, compressible 
\al ic turbulence \cite{us-turb} suggests the existence of two distinct
phases or states of the turbulence, depending on how strong the collisionless
damping is. At last, resonant particles, which were responsible for the wave
damping in the linear Landau process, appear to be trapped in the wave 
potential at large times and the linear theory of Landau damping is invalid. 
Qualitatively, the damping rate oscillates with roughly the bounce frequency 
and gradually decreases to zero until resonant particles phase mix and form a 
plateau on the distribution function. A recently proposed asymptotic theory of 
particle trapping \cite{us-trap} predicts non-trivial nonlinear contributions 
of resonant particles to \na wave dynamics. They are the following.
(1) The power of the KNLS nonlinearity associated with resonant 
particles increases to {\em fourth} order when trapped particles are
accounted for. (2) The effective coupling 
is now proportional to the {\em curvature} of the particle distribution
function (PDF) at the resonant velocity, $f_0''(v_A)$, and not to its slope, 
$f_0'(v_A)$, as in linear theory. (3) The {\em phase-space density} of 
trapped particles is controlled by plasma $\beta$. 
Thus, the structure of quasi-stationary \al ic solitons and discontinuities 
may be modified in $\tau\to\infty$ limit.

The rest of this paper is organized as follows. In Sec.\ \ref{s:dnls} a 
semi-qualitative derivation of the envelope evolution equation for
finite-amplitude \al\ waves, referred to as 
the Derivative Nonlinear Schr\"odinger (DNLS) equation, wave is presented. 
In Sec.\ \ref{s:knls} we generalize it to
include the resonant particles effect (Landau damping) and obtain
the kinetically modified derivative nonlinear Schr\"odinger equation (KNLS).
We discuss numerical solutions for this equation which appear to be the 
{\em \al ic Rotational Discontinuities} that observed {\it in situ} in the
solar wind. We also discuss the energy dissipation time scales for various
types of \na waves. In Sec. \ref{s:turb} the noisy-KNLS model of
strong, compressible magnetohydrodynamic (MHD) \al ic turbulence is discussed. 
In Sec. \ref{s:trap} a self-consistent treatment of 
the nonperturbative effects of particle response
(particle trapping) on the \na wave dynamics is presented.

\section{Nonlinear Alfv\'en Waves: the DNLS Equation} 
\label{s:dnls}

The usual derivation of the evolution equation of the \na waves is based on
the multiple time scale expansion. It assumes that the envelope of an
\al\ wave varies on much longer time-scale, $\tau=(B_0/\tilde B_\bot)^2t\gg t$, 
than the linear wave period, $t=\omega^{-1}=(k_\|v_A)^{-1}$, where 
$B_0$ and $\tilde B$ are the unperturbed magnetic field and its perturbations,
$v_A=B_0/\sqrt{4\pi m_in}$ is the \al\ speed, $m_i$ is the ion mass,
and $\|$ and $\bot$ designate components parallel and perpendicular to the
ambient (unperturbed) magnetic field.
Such a derivation has been performed by different authors and we refer the
reader to the original literature \cite{CohenKulsrud,Kennel-etal}. 
To provide physical insights, we give here 
a transparent, semi-qualitative derivation instead. 

A wave equation of a small amplitude \al\ wave may be written as
\beq
\left(\omega-k_\|v_A\left[1\pm k_\|v_A/2\Omega_i\right]\right)\tilde B_\bot=0 ,
\label{1}
\eeq
where $\Omega_i=eB_0/m_ic$ is the ion-cyclotron frequency. The last term is 
simply weak dispersion, $k_\|v_A/\Omega_i\ll1$, due to a finite Larmor radius.
A ponderomotive force creates density fluctuations, $\delta n$, 
so that the \al\ speed is no longer constant:
\beq
v_A=\frac{B_0}{\sqrt{4\pi m_i(n_0+\delta n)}}
\simeq v_{A0}\left(1-\frac{1}{2}\frac{\delta n}{n_0}\right).
\eeq
We assume that the wave propagates along the magnetic field in $z$-direction
(to the right) and its amplitude is changing gradually, 
$\tilde B_\bot=\tilde B_\bot(z-v_{A0}t,\tau)$. Going back to real-space
representation, this suggests the following replacement:
\beq
\omega\to-iv_{A0}\frac{\partial}{\partial z}+i\frac{\partial}{\partial \tau}\ ,
\qquad k_\|\to-i\frac{\partial}{\partial z}\ .
\eeq
We can neglect the variation of $v_A$ in the small dispersion term as a next
order effect. From Eq.\ (\ref{1}) it is easy to write:
\beq
\frac{\partial}{\partial \tau}\tilde B_\bot-\frac{v_A}{2}
\frac{\partial}{\partial z}\left(\frac{\delta n}{n_0}\tilde B_\bot\right)
\pm i\frac{v_A^2}{2\Omega_i}\frac{\partial^2}{\partial z^2}\tilde B_\bot=0.
\label{dnls1}
\eeq
Hereafter we omit subscript ``$0$'' in $v_A$. The perturbed magnetic field
here is a complex quantity and is defined as 
$\tilde B_\bot=\tilde B_x\pm i\tilde B_y$. The density perturbations are
caused by the ponderomotive force exerted on plasma by the fluctuating
magnetic field of the wave. They are obtained from the continuity, momentum
balance equations, and equation of state which yield the
equation for ion-acoustic waves with source:
\beq
\left(\frac{\partial^2}{\partial t^2}
-c_s^2\frac{\partial^2}{\partial z^2}\right)\frac{\delta n}{n_0}=
-\frac{v_A^2}{2}\frac{\partial^2}{\partial z^2}
\frac{|\tilde B_\bot|^2}{B_0^2} ,
\label{sound}
\eeq
where $c_s^2=\gamma(T_e+T_i)/m_i$ is the ion-acoustic speed, 
$\gamma=3$ is the polytropic constant, and 
$p=n(T_e+T_i)$ and $m_e\ll m_i$ were used in derivation. 
Since, ion-acoustic waves are driven 
by the \al\ wave, the density perturbation is a function of $z-v_At$. Thus 
$\frac{\partial^2}{\partial t^2}\equiv v_A^2\frac{\partial^2}{\partial z^2}$
and Eq.\ (\ref{sound}) yields:
\beq
\frac{\delta n}{n_0}=-\frac{1}{2(1-\beta)}
\frac{|\tilde B_\bot|^2}{B_0^2} ,
\eeq
where $\beta=c_s^2/v_A^2$ is roughly the ratio of plasma pressure to magnetic
pressure. Substituting the density perturbation into the equation for the wave
magnetic field (\ref{dnls1}) and defining $b=\tilde B_\bot/B_0$, we obtain the
DNLS equation:
\beq
\frac{\partial b}{\partial \tau}+\frac{1}{4(1-\beta)}
\frac{\partial}{\partial z}\left(|b|^2 b\right)
\pm i\frac{v_A^2}{2\Omega_i}\frac{\partial^2 b}{\partial z^2}=0.
\label{dnls}
\eeq
This equation describes evolution of planar \na waves propagating 
in one dimension along the ambient magnetic field. It was shown that this
equation is solvable \cite{KaupNewell} and admits soliton and cnoidal 
wave solutions.

A numerical solution of the initial value problem for the DNLS is shown in 
Fig.\ \ref{f1}a. It is natural to assume that nonlinear waves emerge 
from small-amplitude
(linear) ones. Thus the most general class of initial profiles are the
finite-amplitude, periodic waves of different helicities (polarizations).
In the case of Fig.\ \ref{f1}a, an initially linearly polarized, sinusoidal
initial wave profile has been chosen. (The range of dimensionless coordinate,
$0\le\zeta\le1024$ corresponds to the number harmonics used in calculation,
$-512\le k\le512$.) One can see that wave front indeed
steepens at early times, $\tau\sim2\sim400\Omega_i^{-1}$. Later, at
$\tau\sim5$, 
when the width of the front becomes of order $\Omega_i/v_A$ dispersion limits
nonlinear steepening and produces small-scale, oscillatory wave-like
structures with significant small-scale energy component. At later times, the
magnetic field is highly irregular indicating strong \na wave turbulence.
Note that {\em no (quasi-stationary) \al ic discontinuities emerge}.

\section{Nonlinear Alfv\'en Waves with Landau-type Dissipation: 
the KNLS Equation} 
\label{s:knls}

It is known that kinetic effects like Landau damping are absent from the MHD
approximation. To include them into the \na wave model self-consistently, 
a fully kinetic treatment is needed \cite{Rogister}. The results are often 
too complicated and obscuring, however. There are two ways to go around the 
difficulties. First way is to use a hybrid MHD-kinetic approach in which one
performs the {\it ad hoc} calculations the quantities which are 
known to be affected by kinetics (i.e., the density perturbations, $\delta n$) 
and plugs them into the MHD model \cite{MjolhusWyller,Spangler1,Spangler2}.
Second way is to make some self-consistent closures in deriving the MHD
equations to modify them so that to mimic collisionless processes
\cite{HammetPerkins}. Then these new MHD equations may be used to re-derive
the \na wave equation \cite{us-POP1}. Here we again give a simple, 
semi-qualitative derivation. For rigorous analysis, we refer reader to the 
original literature \cite{us-POP1,MjolhusWyller,Spangler1,Spangler2}.

To include kinetic effects, we need to modify the equation for ion-acoustic
waves only, while Eq.\ (\ref{dnls1}) remains unchanged. The equation of damped
ion-acoustic modes may be formally written as 
\beq
\left(\frac{\partial^2}{\partial t^2}+\hat\gamma\frac{\partial}{\partial t}
-c_s^2\frac{\partial^2}{\partial z^2}\right)\frac{\delta n}{n_0}=
-\frac{v_A^2}{2}\frac{\partial^2}{\partial z^2}
\frac{|\tilde B_\bot|^2}{B_0^2} ,
\label{sound-damp}
\eeq
where $\hat\gamma$ is a damping rate. If the damping rate is a function of $k$ 
in Fourier space, $\hat\gamma$ in real space is represented by an operator 
acting on a wave field. We now recall that in $k$-space Landau damping rate is
proportional to $1/|k_\||$. Thus in real space we write
\beq
\hat\gamma=v_A\chi\hat{\cal H}\frac{\partial}{\partial z} ,
\eeq
where $\hat{\cal H}[f]
=\frac{1}{\pi}\int_{-\infty}^\infty\frac{\cal P}{z'-z}f(z'){\rm d}z'$ is the
integral Hilbert operator acting on a function $f$ (${\cal P}$ denotes
principal value integration), which is equal to $(ik_\|/|k_\||)f_k$ in Fourier
space and, thus, ensures correct $k$-scaling of collisionless damping.
Here $\chi$ is a parameter which contains all insignificant physics such as
the structure of a PDF, etc., and depends on the closure chosen. 
In the model of Ref.\ \cite{us-POP1}, $\chi$ has the
physical meaning of a thermoconductivity coefficient.
Again, we assume a traveling wave solution for the density perturbations,
$\delta n=\delta n(z-v_At)$. Than we have
\beq
(1-\beta)\frac{\delta n}{n_0}+\chi\hat{\cal H}\left[\frac{\delta n}{n_0}\right]
=-\frac{1}{2}|b|^2 .
\label{tmp1}
\eeq
To solve this equation we use the identity $\hat{\cal H}\hat{\cal H}=-1$,
which has a physical meaning of {\em time-reversibility} of Landau damping 
\cite{us-POP1}. Acting on Eq.\ (\ref{tmp1}) with $\hat{\cal H}$, excluding
$\hat{\cal H}[\delta n/n_0]$ and substituting into Eq.\ (\ref{dnls1}), we obtain
the KNLS equation: 
\beq
\frac{\partial b}{\partial \tau}+v_A\frac{\partial}{\partial z}
\left(m_1 b|b|^2+m_2 b\hat{\cal H}[|b|^2]\right)
+ i\frac{v_A^2}{2\Omega_i}\frac{\partial^2 b}{\partial z^2}=0 ,
\label{knls}
\eeq
where $m_1=(1-\beta)/4\Delta,\ m_2=-\chi/4\Delta,\ \Delta=(1-\beta)^2+\chi^2$,
are coefficients, which are model-dependent. The rigorous derivation 
\cite{us-POP1} yields
\begin{mathletters}
\label{m}
\begin{eqnarray}
m_1&=&\frac{1}{4}\frac{(1-\beta^*)+\chi^2(1-\beta^*/\gamma)}{
(1-\beta*)^2+\chi^2(1-\beta^*/\gamma)^2 } , \\
m_2&=&-\frac{1}{4}\frac{\chi\beta^*
(\gamma-1)/\gamma}{(1-\beta^*)^2+\chi^2(1-\beta^*/\gamma)^2 } ,
\end{eqnarray}
\end{mathletters}
where $\beta^2=(T_e+T_i)\beta/T_i\simeq(T_e/T_i)\beta$ for $T_i\lesssim T_e$, 
and $\chi$ is the best fit parameter of the closure to the exact kinetic
particle response \cite{HammetPerkins} with a temperature correction, $\chi=
\sqrt{8\beta/\pi\gamma}\exp\{-1/\beta\}(T_e/T_i)^{3/2}\exp\{(T_i-T_e)/2T_i\}$.
The dependence of $m_1$ and $m_2$ vs. $1/\beta$ is drawn in Fig.\ \ref{fig-m}.
Negative sign of $m_2$ indicates damping. The damping of an \al\ wave 
is strongest in warm, isothermal plasmas, $\beta\sim1,\ T_e\sim T_i$.
The KNLS equation straightforwardly generalizes to the case of obliquely
propagating waves. In fact, Eq.\ (\ref{knls}) remains unchanged, but the wave
field is re-defined as follows: $b=(\tilde B_x+i\tilde B_y+B_0\sin\Theta)/B_0$,
where $\Theta$ is the obliquity angle, $\sin\Theta\propto{\bf k\cdot B}_0$.

Clearly, $m_1$ and $m_2$ are functions of $T_e/T_i$ and $\beta$. It can be
shown that if either $T_e/T_i\gg1$, or $\beta\gg1$, or $\beta\ll1$, the Landau
damping of \na\ waves becomes very weak, $m_2\to0$. Then the DNLS nonlinearity 
dominates and $m_1\to(1-\beta)^{-1}$, i.e., changes sign at $\beta\simeq1$.
In this regime, other dissiaption mechanisms (e.g., ion-cyclotron damping
of the linear \al\ wave [carrier]) may become important, see Sec.\ \ref{b}.

\subsection{\al ic discontinuities}

The KNLS equation (\ref{knls}) reduces to the DNLS for $\beta\to0$.
Collisionless damping of \na waves vanishes in this case. The wave dynamics is
thus decoupled from resonant particle effects and \al ic discontinuities do
not emerge (unless one artificially plugs in diffusion, finite plasma
conduction, or other collisional effects, as done in some simulations).
For finite $\beta$'s, the \na wave evolution is affected by resonant 
particles and is described by the second nonlinear, integral term in 
Eq.\ (\ref{knls}). Unlike the DNLS equation, 
the KNLS equation is not integrable because of dissipation.
Numerical solution of this integro-differential equation for the same initial
conditions as in Fig.\ \ref{f1}a and $\beta=1$, reveals different wave dynamics,
as shown in Fig.\ \ref{f1}b. Instead of irregular, fluctuating wave fields,
localized, quasi-stationary waveforms (\al ic discontinuities) are seen to 
form very rapidly, within $\tau\lesssim5=10^3\Omega_i^{-1}$. There are 
three parameters (in addition to $\beta$ and $T_e/T_i$) which control the 
wave dynamics as well as specify the type of discontinuity which results.
These are (i) the wave helicity (i.e., the polarization type), (ii) the 
obliquity angle $\Theta$, and the angle between the polarization plane (for 
nearly linear polarizations) and ${\bf k}$--${\bf B}_0$ plane.

There are three general types of {\em quasi-stationary} \al ic discontinuities,
discriminated by the obliquity and phase jump (i.e., angle though which the 
magnetic field vector rotates). First, there is a wide class of the
{\em arc-type} rotational discontinuities which emerge in {\em oblique} 
propagation, $\Theta\gtrsim10^\circ$, independent of the wave helicity
(polarization). They are characterized by an arc-type shape diagram in the 
hodograph (the $b_x$--$b_y$ diagram), as shown in the ``snap-shot'', Fig.\
\ref{fig-arc}. The wave phase jump through the discontinuity is 
$\Delta\phi<\pi$. Second, there is a class of the 
{\em S-type} directional discontinuities which emerge in {\em quasi-parallel}
propagation, $\Theta\sim0^\circ$, only from linearly and almost linearly 
polarized initial perturbations. They are characterized by a remarkable 
S-shaped hodograph, as shown in Fig.\ \ref{fig-S}. 
They are called directional (not rotational) discontinuities
because the phase jump is accompanied by moderate amplitude variation.
The wave phase changes by $\Delta\phi=\pi$ through the the discontinuity.
Third, there is a narrow class of arc-type rotational discontinuities
propagating parallel to the magnetic field. They emerge from waves with
elliptical polarization only. The wave phase change is $\Delta\phi=\pi$.
Parallel propagating circularly polarized waves (helicity equals to unity)
do not evolve to discontinuities and are decoupled from dissipation.

\subsection{Comparison with observations \label{b}}

There are two regimes of the \al\ wave evolution, depending on the values of
$T_e/T_i$ and $\beta$. Although there is no sharp boundary between the regimes,
a {\em qualitative} insight may be gained from Fig.\ \ref{fig-turb}, where the 
curve $|m_1|=|m_2|$ is plotted. Kinetic damping is important for $T_e/T_i$ and 
$\beta$ from the region labeled ``hydrodynamic''. There $|m_1|<|m_2|$ and 
\al ic rotational discontinuities rapidly form. We emphasize the remarkably 
good correspondence of the KNLS solutions and the solar wind observational
data \cite{LeppingBehannon,Neugebauer,Tsurutani}. Recent $2\frac{1}{2}$ 
hybrid code simulations are also in excellent agreement with the results of 
the KNLS theory \cite{code1,code2} and thus support the idea that the mechanism
of formation of \al ic discontinuities is the combined effect of
resonant wave-particle interactions and nonlinear wave steepening.

In the opposite case, $|m_1|>>|m_2|$ (labeled ``bursty'' in Fig.\ 
\ref{fig-turb}), the Landau damping is weak and may be neglected compared 
to the ion-cyclotron one. Nonlinear wave steepening cascades wave energy 
to the scales $k\sim max\{|m_1|,|m_2|\}|b|^2\Omega_i/v_A$, as estimated 
from Eq.\ (\ref{knls}).
Thus, the number of particles being in the cyclotron resonance,
$k(v_A-v)-\Omega_i=0$, may be large if $m_2\to0,\ \beta\simeq1$. 
Nonlinear dynamics in this case is governed by the so called DNLS-Burgers 
equation \cite{DNLSB}. This theory explains well the emergence and evolution
of \al\ shock trains (also referred to as ``shocklets'') \cite{sh-train},
which were observed upstream of the bow-shocks of planets and comets (see
e.g., Ref.\ \cite{shocklets}).

\subsection{Energy dissipation}

Resonant particle effects make the otherwise almost non-dissipative 
\al\ waves to damp. Fig.\ \ref{fig-E} shows temporal evolution of magnetic
energy associated with different structures (\al ic discontinuities).
Clearly, the energy dissipation rate in the case of quasi-parallel propagation
is sensitive to the initial polarization, i.e., helicity, which is, roughly, a
measure of asymmetry of a spectrum between $+k$ and $-k$ harmonics.
Crude estimates \cite{us-POP2} of the typical damping time are: (i) for 
linearly polarized waves (zero helicity), e.g., the S-type discontinuities, 
$\tau_{\rm lin}\propto |m_2|^{-1}$ and the damping is algebraic,
$|b|^2\propto\tau^{-1/2}$, rather than exponential, (ii) for elliptically 
polarized waves $\tau_{\rm ell}\propto |m_2\Delta_s|^{-1}$, where 
$\Delta_s\sim\sum\left(|b_{+k}|^2-|b_{-k}|^2\right)$ is a measure of spectrum 
asymmetry, and (iii) for circularly polarized waves (helicity unity) 
$\tau_{\rm cir}\to\infty$ as they do not damp. In the case of oblique
propagation, the ambient field enters the damping rate and for all wave 
helicities $\tau_{\rm obl}\propto |m_2\sin^2\Theta|^{-1}$. From Fig.\ 
\ref{fig-E} one sees that the a typical time-scale of the wave damping is 
$\tau\lesssim100\sim10^4\Omega_i^{-1}$, which is (roughly) a few hours for the
solar wind conditions.

Given the spectrum of magnetic field energy perturbations, 
$E_k=(|b|^2)_k$, the instantaneous rate of
energy change is calculated {\em exactly} \cite{us-POP2,MjolhusWyller2} 
from Eq.\ (\ref{knls}) to yield:
\beq
\frac{\partial E}{\partial\tau}
=m_2\int_{-\infty}^\infty\left|E_{k_\|}\right|^2\,|k_\||\,{\rm d}k_\|\ .
\eeq
Note that this equation is equally applicable to the single wave case as well as
to the compressible \al ic turbulence in general, 
provided the turbulence spectrum is known.

Typical dissipation timescale is several hours for the solar wind conditions.
Hence, collisionless dissipation of nonlinear \al\ waves a plausible 
(and natural) mechanism of the solar wind heating within one astronomical unit.

\section{Compressible Turbulence of Nonlinear Alfv\'en Waves}
\label{s:turb}

There are two approaches in the studies of turbulence, 
cascade and structure-based. The cascade-type theories 
usually treat turbulence as a ``soup of eddies'', --- a collection of plasma
excitations with random phases which interact with each other (usually on
matching scales) and produce a cascade of energy. The structure-based-type 
approach, one of the examples of which is the noisy Burgers model in
hydrodynamics, is based on the study of nonlinear evolution equations with
external noise drive. It thus studies the turbulence of coherent structures,
such as shocks, solitons, nonlinear (cnoidal) waves, etc., generated
and interacting with each other randomly. Such a process may be referred to as 
``coherent cascade'', meaning that modes (eddies) at all scales interact 
coherently (in phase) and, thus, experience stronger interactions due to the
cumulative effect. Its is known that the first approach is more suitable for 
studies of weak turbulence and the second one better describes the strong, 
highly nonlinear turbulence.

The analytical study of the structure-based noisy-KNLS model 
as a generic model of collisionless, large-amplitude, 
compressible MHD (\al ic) turbulence  is presented 
in Ref.\ \cite{us-turb}. Stationarity is maintained via the balance
of noise and dissipative nonlinearity.
The Fourier-transformed KNLS equation (\ref{knls}) 
with noise source reads
\begin{eqnarray}
& &\left(-i\omega+iv_0k+i\mu_0k^2\right)b_{k\atop\omega}
+i\lambda k\!\!\sum_{k', k''\atop\omega', \omega''}\!\!
b_{k'\atop\omega'}b_{k''\atop\omega''}
b_{k-k'-k''\atop\omega-\omega'-\omega''}
\nonumber\\ & &{ } \quad\times
\left[m_1+im_2\textrm{sgn}(k-k')\right]=f_{k\atop\omega} ,
\end{eqnarray}
where $v_0$ and $\mu_0$ are the bare phase velocity (it is necessary for
a self-consistent renormalization analysis) and dispersion, $\lambda=1$ is 
the standard perturbation parameter, and the function $\textrm{sgn}(x)=x/|x|$.
Here we omitted the subscript $\|$ by k.
The one loop renormalization group analysis of the problem with zero mean,
$\delta$-correlated noise with white $k$-spectrum yields complex-valued 
renormalized coefficients $v_{\rm turb},\ \mu_{\rm turb}$. 
The `Re' and `Im' parts of $\mu_{\rm turb}$ are simply turbulent 
dispersion and turbulent viscosity. The real part of $v_{\rm turb}$ 
may be interpreted as wave momentum loss via interactions with resonant 
particles and its imaginary part manifests fast, exponential damping due to
phase mixing. Further analysis shows the existence of two different phases of
turbulence. The bifurcation occurs at the point 
\beq
|m_1/m_2|\simeq1.
\eeq 
If $|m_1/m_2|<1$, Landau dissipation is efficient to sink all the injected
energy during the ``coherent cascade'' so that the hydrodynamic 
($\omega\to0,\ k\to0$) regime with no sharp fronts realizes.
In this regime, the strongly interacting \al ic discontinuities dominate in 
the turbulence. In the opposite case, nonlinearity overcomes damping and no 
stationary, hydrodynamic turbulence is predicted. Small-scale, bursty 
turbulence is expected in this regime. Given the weak cyclotron damping, such
a regime corresponds to the \al ic shocklet turbulence.
Recalling that $m_1$ and $m_2$ are functions of $\beta$ and $T_e/T_i$,
a ``diagram of state'' can be drawn, as shown in Fig.\ \ref{fig-turb}.

\section{Asymptotic, $\tau\to\infty$, Dynamics of Dissipative Nonlinear
Alfv\'en waves} 
\label{s:trap} 

The linear Landau damping theory used in derivation of the KNLS assumes a
time-independent (Maxwellian) PDF. It is clear that, for a finite-amplitude 
wave, particles which are near resonance with the wave, $v\simeq v_A$, 
will be trapped by the wave potential $U(z)\equiv\tilde B_\bot^2/8\pi n_0$
because their kinetic energy (measured in the wave frame), 
$\frac{1}{2}m(v-v_A)^2$, is less than the potential barrier, 
$|U_m|=max|U(z)|$, as in Fig.\ \ref{fig-trap}. 
Such particles experience reflections at turning points $z_1$ and $z_2$
exchanging energy with the wave and significantly modify the PDF near 
resonance. The damping rate of a wave oscillates in time with gradually
decreasing amplitude until phase mixing results in flattening of 
the PDF (for resonant velocities) and formation of a {\em plateau}; then the
damping rate vanishes \cite{Shapiro}. 
Thus, the {\em linear} calculation of Landau dissipation, 
while correct for times short compared to the typical  bounce (trapping) 
time, $\tau\ll\tau_{tr}\simeq(k_\|\sqrt{U/m_i})^{-1}\simeq(k_\|v_A|b|)^{-1}$, 
fails for quasi-stationary waveforms (\al ic discontinuities) on times 
$\tau\gtrsim\tau_{NL}\gg\tau_{tr}$ 
[$\tau^{-1}_{NL}\simeq m_1 k_\|v_A (\tilde B_\bot^2/B_0^2)$ 
is the typical nonlinear 
wave profile evolution time]. Hence, Landau dissipation should be
calculated {\em non-perturbatively} to determine the resonant particle 
response to the nonlinear wave.

The nonlinear Landau damping problem is, in general, not 
{\em analytically} tractable, as it requires explicit expressions for 
{\em all} particle trajectories as a function of initial position and time. 
Such trajectories cannot be explicitly calculated for a potential
of {\em arbitrary shape}. Usually, a full particle simulation is required. 
In some cases, it is useful to approximate the wave profile shape by a 
simple analytic expression which may be assumed to persist, while the
wave amplitude varies, to calculate the trajectories \cite{ONeil}. 
This is not the case for our problem because 
nonlinear Landau damping controls the profile of the \al\ wave.

A nonperturbative, self-consistent theory of nonlinear wave-particle 
interactions was constructed in the asymptotic limit, $\tau\to\infty$,
applying the virial theorem to determine the PDF. 
The generalized KNLS again is Eq.\ (\ref{dnls1}) with 
$\delta n=\delta n_{NR}+\delta n_R$, where $\delta n_{NR}$ and 
$\delta n_R\propto\chi\hat{\cal K}[U(z)]$
are the non-resonant (bulk) and resonant particle responses, respectively.
Here $\hat{\cal K}$ is a new kinetic operator
(which replaces $\hat{\cal H}$) acting of the wave field which
contains all the information about resonant (trapped) particle trajectories.
It is interesting that the very possibility of writing the generalized KNLS
equation in this form relies on the intrinsic {\em time
reversibility} of the Vlasov equation, linear or nonlinear:
$\hat{\cal K}\hat{\cal K}=-1$, see Ref.\ \cite{us-POP1}. 

The resonant particle response is calculated using {\em Liouville's theorem}
(``the PDF is constant along particle trajectories''):
\beq
f(v,z,t)=f_0\left(v_\pm\left(E,z_0^\pm\right)\right) ,
\eeq
where $z_0^\pm=z_0^\pm\left(z,t,E;U(z)\right)$ is the 
{\em initial} coordinate of a particle of total energy $E$ which at time 
$t$ is at the point $z$ and has a velocity $v_\pm(E,z)$. 
By definition 
$\delta n_R=\int_{\Delta v_{res}}{\rm d}v\left(f-f_0^{(t=0)}\right)$, so that
\beq
\delta n_R=\sum_{(\pm)}\int_{\Delta E_{res}}\!\!\!\!\!
\frac{f_0\!\left(v_\pm\!\left(E,z_0^\pm\right)\right)%
-f_0\!\left(v_\pm\!\left(E,z\right)\right)}%
{\sqrt{2m_i\left(E-U(z)\right)}}\; {\rm d} E .
\label{dnR}
\eeq
Here the sum is over particles moving to the right (+) and to the left (--),
as in Fig.\ \ref{fig-trap}. The integration is over the resonant
(negative) energies of trapped particles, $U_m\le E \le 0$ with $U_m$
being the amplitude of the potential. In the short-time limit,
$\tau\to\infty)$, Eq.\ (\ref{dnR}) reduces to the KNLS case with 
$\hat{\cal K}=\hat{\cal H}$ and $\chi=\pi v_A^2f'_0(v_A)/m_in_0$.

To treat the $\tau\to\infty$ limit we employ the {\em virial theorem}, which
relates the average kinetic and potential energies of 
particles trapped in an adiabatically changing potential:
\beq
2\langle K(z)\rangle=n\langle\tilde U(z)\rangle .
\label{virial}
\eeq
Here $\tilde U(z)=U(z)-U_m$ is a {\em homogeneous} function of its argument 
of order $n$, i.e., $\tilde U(az)=a^n\tilde U(z)$.
Direct integration of Eq.\ (\ref{dnR}) yields
\begin{eqnarray}
\biggl.\langle\delta n_R\rangle\biggr|_{\tau\to\infty}
&\simeq& f_0''(v_A)
\sqrt{\frac{2}{m^3_i}}\sqrt{|U(z)|}
\biggl[\frac{n}{n+2}
\nonumber\\
& &{ }\times\left(|U_m|-|U(z)|\right)-\frac{2}{3(n+2)}|U(z)|\biggr] .
\label{large-t}
\end{eqnarray}
Note that the term $\propto f_0'(v_A)$ vanishes identically, hence
the damping is absent. Since 
$\langle\hat{\cal K}\rangle\langle\hat{\cal K}\rangle\not
=\hat{\cal K}\hat{\cal K}=-1$,
we can only estimate the coupling 
constant to be $\chi\sim f_0''(v_A)\left[v_A^3/n_0\right]\sqrt{2/m^3_i}$.
The index $n$ is formally not defined for an arbitrary potential, but it can
be estimated numerically, given the wave profile \cite{us-trap}.

Clearly, in the asymptotic limit, $\tau\gg\tau_{tr}$, the damping rate vanishes
due to phase mixing. Nevertheless, the resonant particles still contribute
the wave dynamics, in that 
\begin{equation}
\langle\delta n_R\rangle\sim f_0''(v_A) |b|^3 ,
\end{equation}
thus determining a new nonlinear wave equation.

Simple Bernstein-Green-Kruskal type analysis shows that the `height' of the
asymptotic plateau on the PDF depends on the plasma parameters through
$m_1$ and $m_2$ \cite{us-trap}. The difference between the plateau and 
the initial, Maxwellian PDF evaluated at $v_A$ is 
\beq
f_{\rm plat}-\left.f_{\rm Maxw}\right|_{v_A}^{ }=\frac{m_1}{m_2}\frac{|b|}{v_A}.
\eeq
Thus we conclude that there is under-population of trapped particles
(void on the PDF, $f_{\rm plat}<f_{\rm Maxw}$) at low $\beta$'s and 
over-population (bump, $f_{\rm plat}>f_{\rm Maxw}$) at high $\beta$'s.

Finally, assuming weak damping of a \na wave (e.g., ion-cyclotron damping),
one can see \cite{us-trap} from the conservation of the parallel adiabatic 
invariant $J=\oint p_\|{\rm d} z$ that trapped particles will condense near the 
bottom of the wave potential: $\Delta v_\|\propto |b|$, 
forming a clump on the PDF at $v\simeq v_A$.

\section{Acknowledgments}

It is a pleasure to thank Patrick Diamond, Valentin Shevchenko, 
Marshall Rosenbluth, and Vitaly Galinsky for valuable contributions to this
work. We also thank Roald Sagdeev, Shadia Habbal, Vitaly Shapiro, 
Ramesh Narayan, Eliot Quataert, and Sally Ride for interesting discussions.

\begin{figure}
\caption{ Wave profile evolution of a linearly polarized wave with a
sinusoidal initial profile propagating parallel to the magnetic field 
for $T_e\simeq T_i$ and  $\beta=0$ (a) and $\beta=1$ (b).}
\label{f1}
\end{figure}
\begin{figure}
\caption{The coefficients $M_1=2m_1$ and $M_2=2m_2$ vs. $1/\beta$ 
for $T_e=T_i$.}
\label{fig-m}
\end{figure}
%
%
\begin{figure}
\caption{The arc-type rotational discontinuity propagating obliquely to the
magnetic field, $\Theta\sim30^\circ$, with $\Delta\phi<\pi$,
(a) - amplitude and phase profiles,
(b) - hodograph.}
\label{fig-arc}
\end{figure}
\begin{figure}
\caption{Same as Fig.\ \ref{fig-arc} for the S-type directional discontinuity
propagating parallel to the field.}
\label{fig-S}
\end{figure}
\begin{figure}
\caption{Wave energy evolution.}
\label{fig-E}
\end{figure}
\begin{figure}
\caption{The $\beta$-$T_e/T_i$-diagram of state.
The region inside the curve corresponds to highly damped turbulence. No
steep fronts appear. There are steep fronts in the region outside the curve. }
\label{fig-turb}
\end{figure}
\begin{figure}
\caption{Trapping potential.}
\label{fig-trap}
\end{figure}
%
\newpage
NOTE TO EDITOR:\\
ALL FIGURES SHOULD BE SCALED TO THE SINGLE COLUMN FORMAT.\\ .\\ . \\ . \\ .
\hspace*{-0.5in} Fig.~1:\\ ~~~\psfig{file=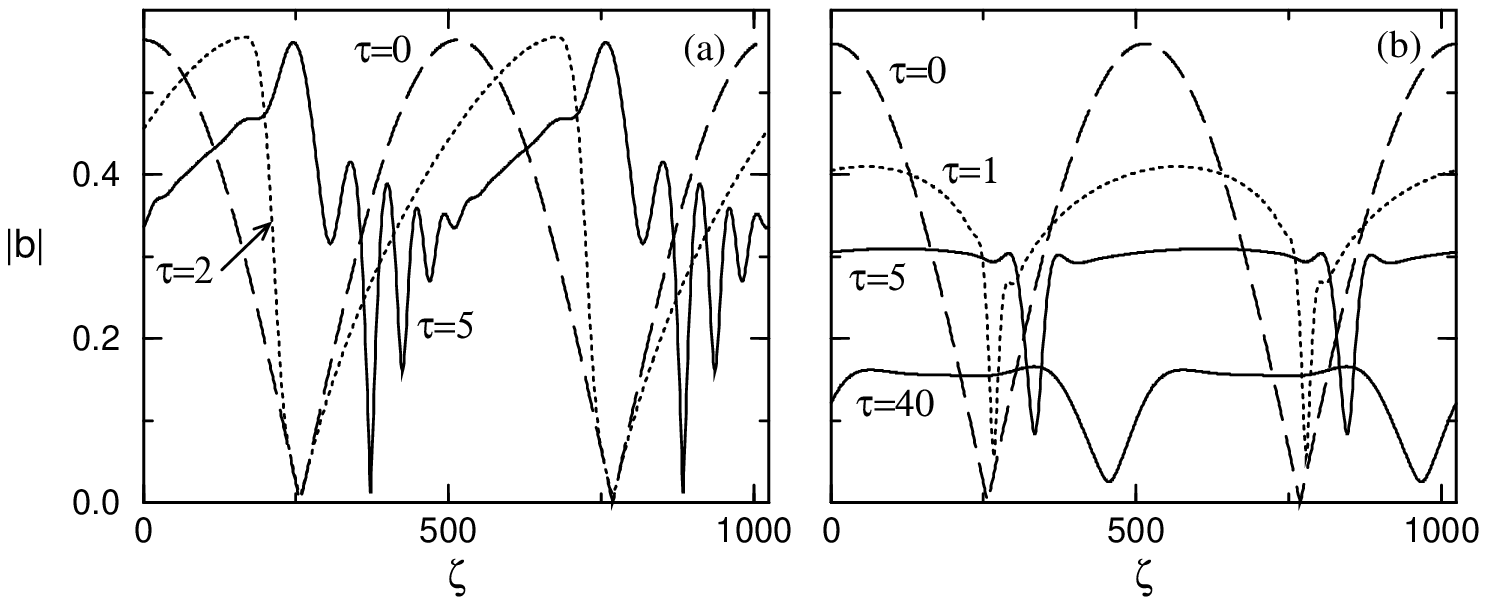}\vspace*{-4in}
\vskip-12.5in\hspace*{-0.5in}Fig.~2:~~~\psfig{file=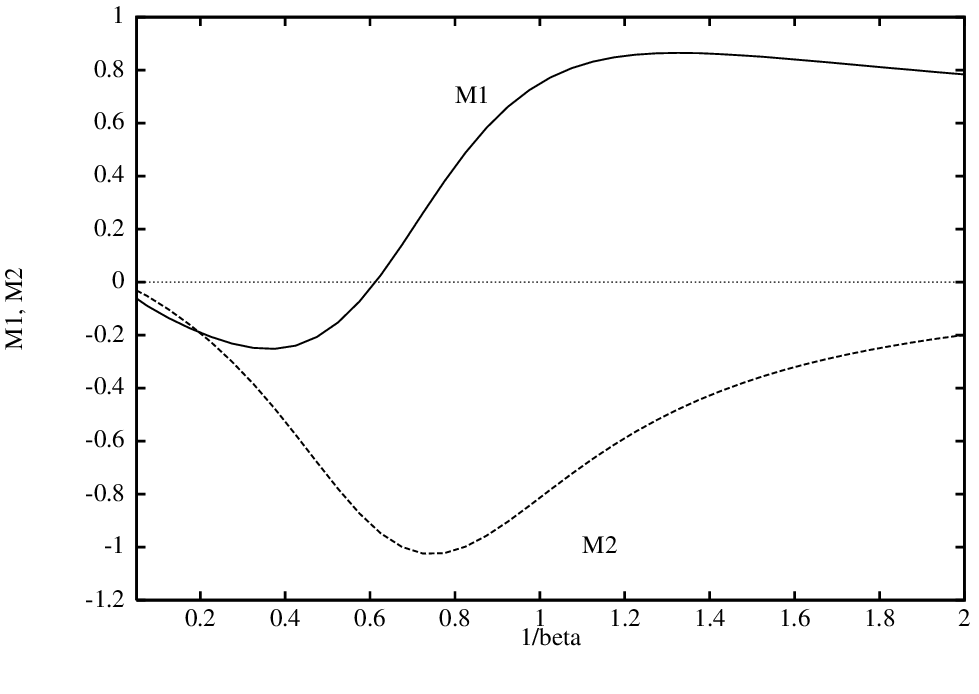}
\thispagestyle{empty}
\newpage\ ~
\newpage
\hspace*{-1in}\psfig{file=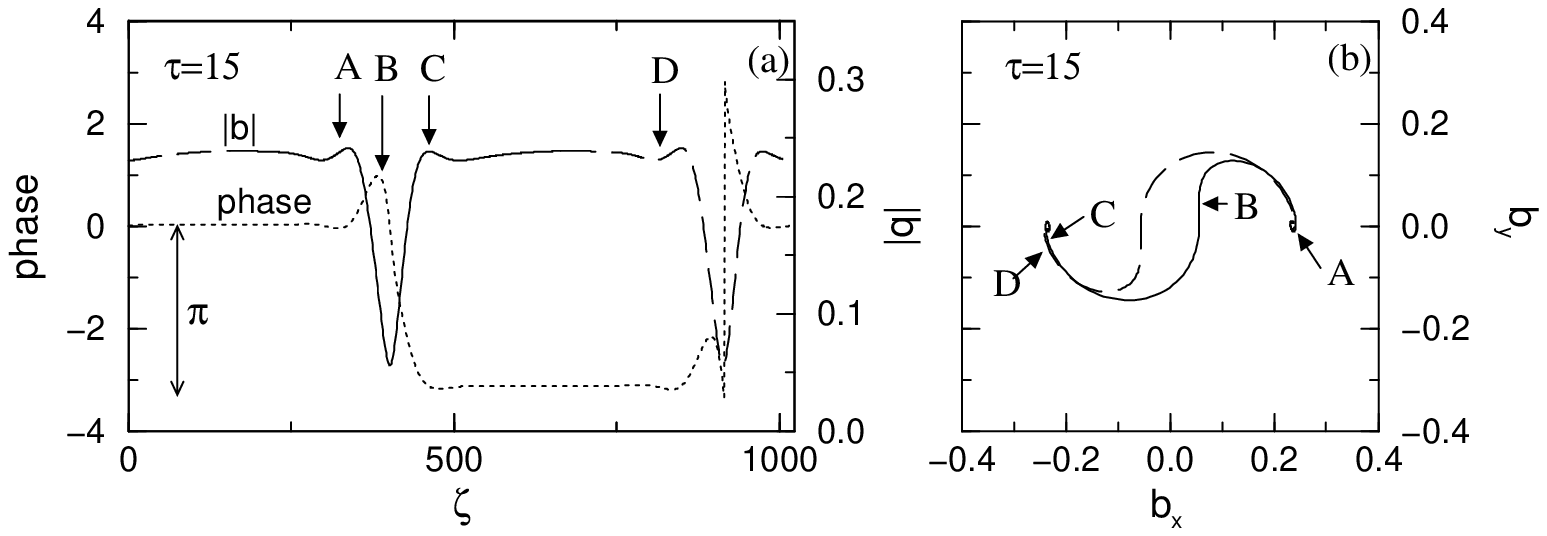}\vspace*{-8in} Fig.~3\vskip0.5in
\hspace*{-1in}\psfig{file=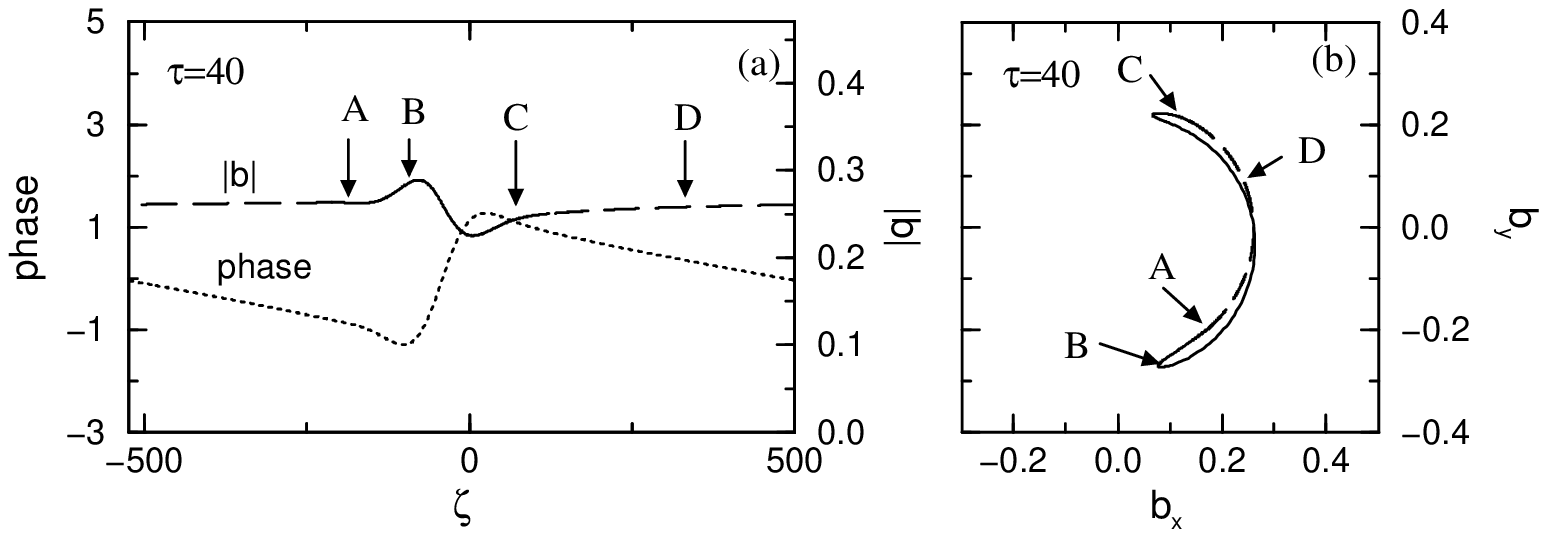}\vspace*{-8in} Fig.~4\vskip0.5in
\hspace*{-1in}\psfig{file=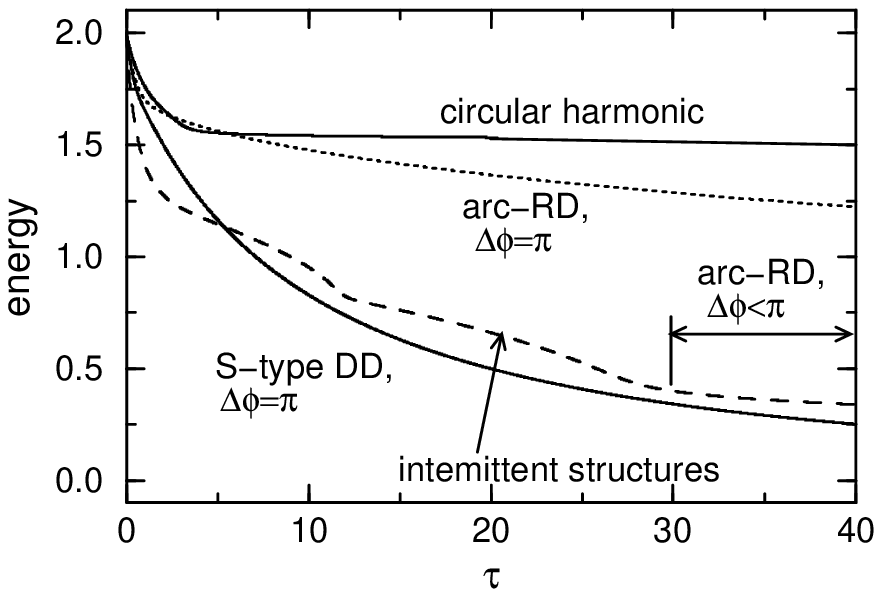}\vspace*{-8in} Fig.~5\vskip0.5in
\thispagestyle{empty}
\newpage\ ~
\newpage
\vspace*{1in}
Fig.~6:\\ 
\sf
\large\sf
\vspace*{0.5in}
\hspace*{3.8in}
\makebox(0,0){\psfig{file=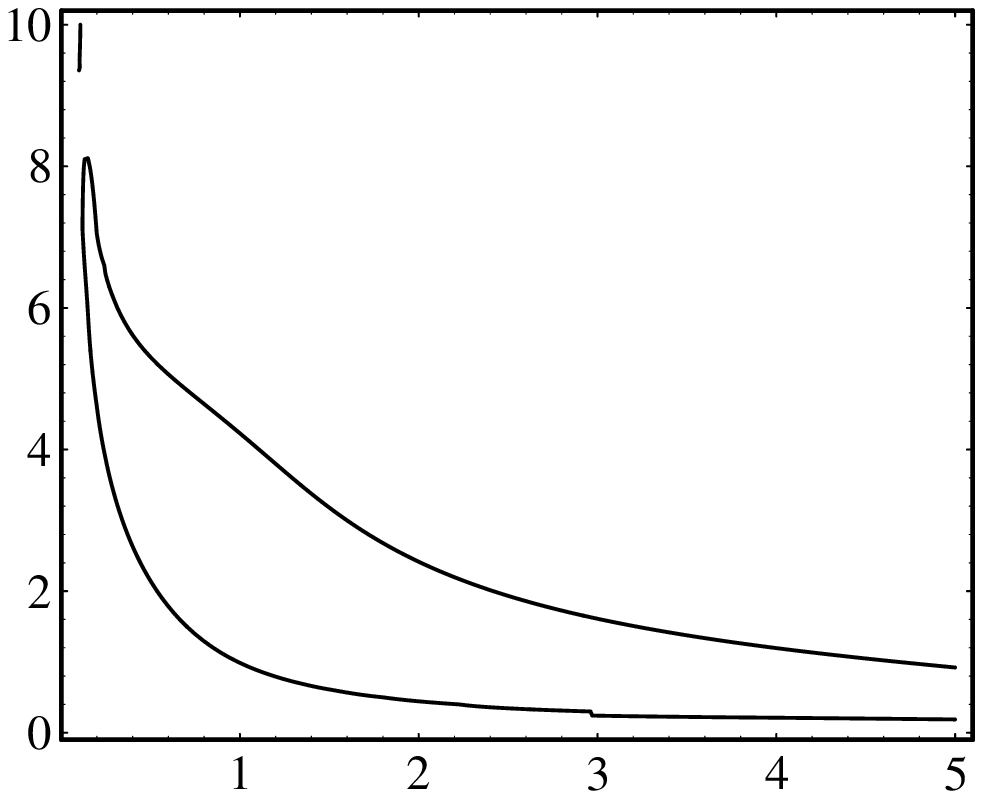,height=8in,width=7in}}
\put(-25,51){$\beta$}
\put(-194,180){$\displaystyle{\frac{T_e}{T_i}}$}
\put(-105,95){hydrodynamic}
\put(-155,85){bursty}
\put(-30,140){bursty}
\vspace*{2in}
\vskip-2in 
\rm Fig.~7:\\ 
\hspace*{1in}\psfig{file=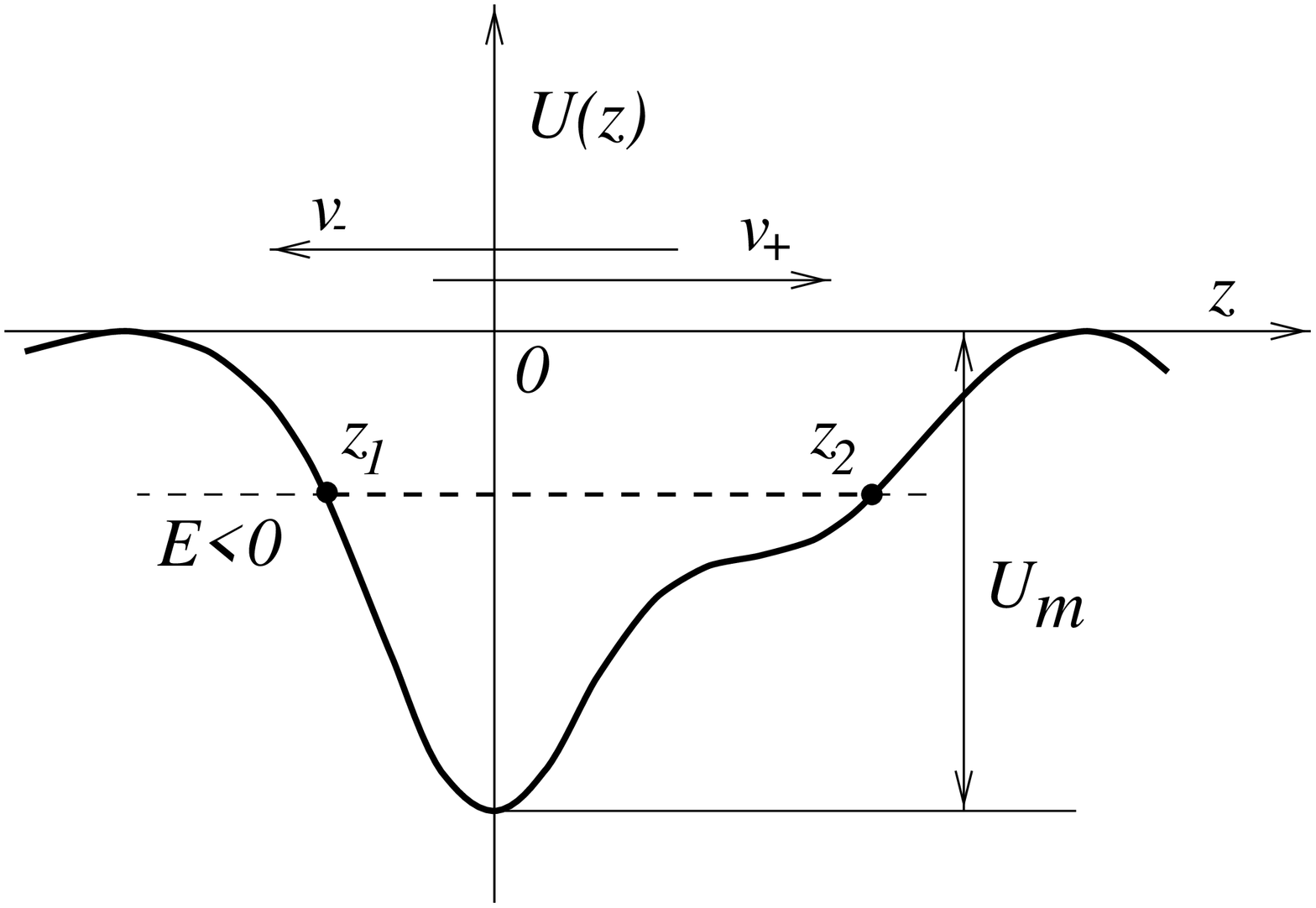,width=5in,angle=-90}
\thispagestyle{empty}
\end{document}